\documentclass[aps,pra,reprint,superscriptaddress,nofootinbib]{revtex4-2}

\usepackage[utf8]{inputenc}
\usepackage[T1]{fontenc}
\usepackage{lmodern}
\usepackage[english]{babel}
\usepackage{microtype}
\usepackage{hyperref}
\usepackage{verbatim}
\usepackage{amsmath,amssymb}
\usepackage{graphicx}
\usepackage{csquotes}
\usepackage{xcolor}

\hypersetup{
  colorlinks=true,
  linkcolor=blue,
  citecolor=blue,
  urlcolor=blue
}

\newcommand\hs[1]{HS$^3$}

\begin{document}

\title{\hs3: A Descriptive, Interoperable Serialization Standard\\
for Statistical Models in High-Energy Physics}

\author{Carsten Burgard}
\affiliation{Department of Physics, TU Dortmund University, Dortmund, Germany}
\affiliation{Institut für Experimentalphysik, Universität Hamburg, 22761 Hamburg, Germany}
\author{Oliver Schulz}
\affiliation{Max Planck Institute for Physics}
\author{Giordon Stark}
\affiliation{Santa Cruz Institute for Particle Physics, UC Santa Cruz}
\author{Jonas Rembser}
\affiliation{CERN, Geneva}
\author{Simon Cello}
\affiliation{Department of Physics, TU Dortmund University, Dortmund, Germany}
\author{Cornelius Grunwald}
\affiliation{Department of Physics, TU Dortmund University, Dortmund, Germany}

\date{\today}

\begin{abstract}
Statistical models in high-energy physics (HEP) formally encode the mathematical relationship between
observed data, physics parameters of interest, and experimental and theoretical 
uncertainties. Likelihood-based inference is the central tool to analyze precision measurements and, for example, perform effective field theory (EFT) fits and cross-analysis combinations. Due to this, there is an increasingly urgent need for 
machine-readable, descriptive, and portable model representations. Existing formats such as \texttt{ROOT} workspaces, pyhf \texttt{JSON}, and CMS 
DataCards each provide valuable capabilities but remain tied to specific software stacks and offer no universal standard for exchange, validation, or long-term preservation.

We introduce \hs3, the \emph{High-Energy Physics Statistics Serialization Standard}, an implementation-agnostic, human-readable, and extensible serialization format for statistical models in high-energy physics. \hs3 is designed such that new statistical constructs can be incorporated through backward-compatible extensions, while inference procedures and implementation-specific execution details remain the responsibility of downstream frameworks. \hs3 represents likelihoods as computational graphs composed of named 
distributions, functions, datasets, domains, and analysis prescriptions. It supports both 
binned and unbinned likelihoods and hierarchical composite models. \hs3 is convertible from and to \texttt{ROOT}/\texttt{RooFit}, and is a superset of pyhf. We describe the design 
principles, structure, and semantics of \hs3 and summarize existing implementations in C++, 
Python, and Julia. We also show some early applications to public likelihoods on HEPData, 
cross-framework validation, and cross-domain reproducibility efforts. \hs3 
establishes a foundation for FAIR (Findable, Accessible, Interoperable, Reusable), long-lived statistical models at the LHC 
and beyond. The standard is meant to cover the needs of the scientific community in general, and to evolve over time to be applicable in a wide range of domains.
\end{abstract}

\maketitle

\section{Introduction}

Likelihood-based statistical models have become the primary scientific artifacts of modern 
high-energy physics. They provide the most complete description of how observed data 
depend on underlying physical parameters and the associated experimental and theoretical 
uncertainties. The focus of LHC physics has evolved from discovery to precision measurements, dominated by systematics, and global interpretations. This means that the ability to publish, exchange, and preserve full likelihood models is no longer just desirable, but has become vital for scientific reproducibility, reinterpretation, and long-term scientific value.

Despite this importance, the modeling ecosystem has remained fragmented. \texttt{ROOT} 
workspaces, pyhf \texttt{JSON} files, CMS DataCards, and emerging Python and Julia toolkits 
offer powerful capabilities, but none provide a general, software-independent, long-term stable on-disk format. Conversions between the above formats are often brittle, validation is 
difficult, and long-term preservation is limited by tight coupling to specific software 
stacks. These challenges have motivated the development of a descriptive, interoperable, 
and durable representation of statistical models.

\hs3, the \emph{High-Energy Physics Statistics Serialization Standard}, addresses this 
need by defining a human-readable, computational-graph-based language\footnote{Some examples are shown in the appendix.} for probability 
distributions, functions, datasets, domains, and analysis prescriptions. \hs3 is designed to 
be expressive enough to represent the full range of models used at the LHC — both binned and 
unbinned — while remaining agnostic to inference engines and algorithms. Implementations 
in \texttt{ROOT}, Python, and Julia will enable transparent exchange of models across programming languages and tool chains, and so decouple the computing environment of likelihood publishers and consumers -- an ideal basis for FAIR open-data practices (Findable, Accessible, Interoperable, Reusable). Cross-language implementations also facilitate independent validation of serialized models across frameworks, helping to identify implementation-specific issues, isolate model-definition errors, and improve confidence in both the standard and its associated tooling.

In the following we present the motivation, design principles, structure and semantics of \hs3, the state of its implementations, and some early applications. We aim to establish  \hs3 as a stable and extensible foundation for long-term preservation and interoperable use  of statistical models in high-energy physics in particular, but also for the scientific community in general.

\section{Background and Motivation}

\subsection{Likelihoods as primary scientific artifacts}

The publication of statistical models in high-energy physics (HEP) has long been 
recognized as essential for the preservation, reinterpretation, and combination of 
experimental results. Arguably, HEP has been a leading science domain in putting this principle into practice. Already in the seminal Workshop on Confidence Limits in 2000, 
a consensus was reached that likelihood functions should be published alongside 
experimental measurements \cite{workshop2k}, as they represent the most complete and unambiguous 
description of the underlying statistical information. Two and a half decades later, this is being done routinely -- and yet, the primary goal has still not been fully realized in a consistent and interoperable manner.

The importance of publishing full likelihoods grew dramatically with the transition 
from the LEP era to the Large Hadron Collider (LHC). While LEP had already embraced 
computerized workflows, its statistical models and data volumes were modest by 
contemporary standards. The LHC, by contrast, confronted physicists and engineers with 
an unprecedented data challenge: discovering the Higgs boson required not only new 
experimental machinery but also an entirely new scale of data analysis. As the 
likelihood function became the central tool for extracting physics from data, the need 
for robust, transparent, and reproducible statistical modeling grew correspondingly.

\subsection{Existing practice and technological limitations}

\texttt{ROOT} emerged early as the dominant data-processing environment at the LHC \cite{ROOT_NIMA_1997}, but initially 
lacked a flexible, fully featured modeling and fitting engine. This capability was 
supplied by {\texttt{RooFit}} \cite{roofit}, originally developed for the BaBar experiment. Unlike the 
predominantly binned analyses used in ATLAS and CMS, many measurements in the 
$B$-physics community rely on analytical functions and unbinned likelihoods, requiring a 
modeling language capable of supporting a wide variety of functional forms. \texttt{RooFit} was 
designed precisely with this diversity in mind, representing statistical models as 
composable computational objects.

As the Higgs discovery approached, the ability to combine information from multiple 
channels became critical \cite{atlashiggs,cmshiggs,higgsmasscomb,higgscouplingcomb}. This could only be achieved at the level of the likelihood 
function. To make this possible, {\texttt{RooWorkspace}} was introduced as a mechanism 
to serialize, store, and combine full statistical models. In practice, the complete specification of a statistical inference problem was formalized through additional descriptive objects included in the \texttt{RooWorkspace} alongside the pdf and data, most notably \texttt{RooFit::ModelConfig}, defining parameters of interest, nuisance parameters, observables, and snapshots.  The design of the \texttt{RooWorkspace} is based on 
a simple but powerful paradigm: any likelihood model can be represented as a \emph{computational graph}
whose nodes are probability distributions, functions, parameters, and datasets. Practically, this representation can be constructed via explicit instantiation of objects, using high-level frameworks such as \texttt{HistFactory} \cite{hf}, or using the workspace factory language with its compact, math-like syntax. Coupled 
with \texttt{ROOT}'s built-in binary serialization system based on \texttt{ROOT}'s C++ reflection capabilities, it became possible to exchange complex models as reusable artifacts for the first time. For nearly a decade, 
\texttt{RooWorkspace}s served as the de-facto standard for sharing statistical models within and 
between LHC collaborations. Notably, the expressiveness of the \texttt{RooWorkspace} is in principle unlimited, as it places no assumptions on the model structure. 

Despite this success, by the late 2010s several obstacles prevented the publication of 
full statistical models. First, \texttt{RooWorkspace}s are binary files tied to specific \texttt{ROOT} 
versions. Although the \texttt{ROOT} developers invested substantial effort in ensuring backward 
compatibility, the complexity of real analysis workspaces made it difficult to guarantee 
long-term compatibility. Second, complete LHC likelihoods are large and intricate, 
typically comprising dozens of analysis regions  (i.\,e.~statistically independent subsets of the data defined by event selections and associated observables), many samples, and hundreds to thousands of 
nuisance parameters, creating concerns about incorrect external usage. Third, 
final models often undergo late-stage modifications — pruning uncertainties, adjusting 
interpolation schemes, modifying constraint terms, or smoothing statistical 
fluctuations — that do not always map cleanly onto the conceptual equations presented in 
publications. Finally, \texttt{ROOT} remained the only software capable of interpreting a 
\texttt{RooWorkspace}, and no alternative modeling framework of the time offered equivalent 
expressiveness.

\subsection{Emergence of descriptive formats: the \texttt{pyhf} breakthrough}

Around 2018, both ATLAS and CMS had largely converged on the HistFactory modeling 
paradigm \cite{hf}: the ``stacks-of-histograms'' approach formalized in terms of binned templates 
and modifiers. CMS provided a unified statistics frontend through their \texttt{Combine} 
toolkit \cite{combine} based on RooFit, while ATLAS lacked an equivalent interface for a long time, with \texttt{TRExFitter} \cite{trexfitter} emerging only recently as the currently dominant equivalent. Both of these, notably, produce \texttt{RooWorkspace}-based likelihoods and thus essentially serve as an easy-to-configure user-facing frontend to the unlimited complexity of \texttt{RooFit}.

This environment created an opportunity for \texttt{pyhf} \cite{pyhf,pyhf-joss}, a pure-Python implementation 
of the HistFactory formalism, to gain rapid adoption. Its lightweight software stack, 
excellent documentation, and, most importantly, its human-readable \texttt{JSON} model 
representation offered exactly the features needed to overcome longstanding reluctance 
towards likelihood publication. For the first time, complex HEP likelihoods could be 
inspected, validated, edited, and reused without specialized software. The first 
likelihoods published in \texttt{pyhf} \texttt{JSON} format were met with enthusiastic adoption by 
the theory community, and none of the feared problems—such as spurious discoveries caused 
by misuse—materialized. This success demonstrated the feasibility and scientific value of 
publishing statistical models in a descriptive format, effectively marking a cultural 
turning point.

Within ATLAS, the absence of a unified statistics frontend similar to CMS's \texttt{Combine} throughout the 2010s
made \texttt{pyhf} additionally attractive for internal use. Publishing models became a 
matter of convenience: if an analysis was already using \texttt{pyhf}, no conversion or 
validation steps were needed, lowering the barrier for public release even further.

\subsection{Fragmentation, interoperability challenges, and the need for a general standard}

The success of \texttt{pyhf} also highlighted an emerging challenge: fragmentation of the 
modeling ecosystem. LHC likelihoods were now being published in a mixture of formats: \texttt{ROOT} workspaces, \texttt{pyhf} \texttt{JSON} files, and later also CMS DataCards — each tied to different 
software frameworks, languages, and assumptions. Conversions between these formats were 
possible but fragile, and verifying equivalence across toolchains was often impractical.

Simultaneously, the broader HEP community was moving away from a pure \texttt{C++} workflows toward a mix of languages. Python became increasingly popular as the user front-end and lately Julia has been emerging as a promising high-performance alternative for both front-end and back-end code. The proliferation of statistical toolkits -- \texttt{zfit} \cite{zfit}, \texttt{BAT.jl} \cite{bat.jl} \& JuliaHEP tools \cite{julia-in-hep-potential,juliahep-perf,julia-in-hep}, BLUE \cite{BLUE} for 
unfolded results -- reflected the need for flexibility, but further exacerbated the lack of a 
common descriptive format. Cross-experiment combinations (ATLAS–CMS \cite{higgscouplingcomb,higgsmasscomb}, later CMS–LHCb \cite{cmslhcb}) would remain tied to the \texttt{ROOT} ecosystem, as it remained the dominant toolkit, but required increasingly close collaboration, with modeling components (e.g.\ CMS 
\texttt{Combine} extensions such as the cascading minimizer \cite{combine}) complicating standardization efforts.

At the same time, the scientific context was shifting. As the LHC entered Run~3, the early 
expectation of rapid discoveries gave way to a sustained focus on precision measurements, 
systematics-limited analyses, and effective field theory (EFT) interpretations spanning 
multiple processes. These developments greatly increased the need for likelihood-level 
combinations and public availability of complete statistical models.

Finally, long-term preservation concerns became increasingly pressing. If the LHC yields 
no new discoveries during its operational lifetime, its legacy will rest on the ability 
of future generations to reinterpret its measurements. Ensuring this legacy requires data 
formats that remain comprehensible and usable independently of the software stacks that 
created them. This broader shift toward open and FAIR science, encompassing event-level 
data, calibrations, workflows, and metadata, motivates the need for a descriptive and 
machine-independent representation of statistical models as well. \hs3 is one component of 
this larger effort.

\subsection{Summary: the need for a general descriptive standard}

Likelihood functions form the mathematical core of nearly all statistical inference in 
HEP. They encode the full dependence of observed data on underlying parameters, including 
physics parameters of interest and experimental or theoretical nuisance parameters. For 
reinterpretation, EFT fits, precision measurements, and cross-process combinations, the 
likelihood is the primary scientific artifact. As systematic uncertainties have become the 
dominant limitation in many LHC Run~2 and Run~3 analyses, and as the EFT framework has gained 
importance, the need to combine statistical information across analyses, processes, and 
experiments has only intensified. These considerations make the publication of complete 
likelihood models not merely beneficial, but essential.

Existing model formats each offer valuable functionality but fall short of providing a 
general, interoperable solution. \texttt{RooWorkspace}s are powerful but binary and tied to \texttt{ROOT}; 
HistFactory-based formats such as \texttt{pyhf} \texttt{JSON} are descriptive but limited in scope; 
CMS DataCards are human-readable but tied to specific \texttt{C++} implementations; and new 
toolkits lack a common serialization standard. The result is a fragmented ecosystem where 
validation is difficult, combinations across frameworks are error-prone, and long-term 
preservation is not guaranteed.

The technological evolution of HEP, the increasing importance of EFT and global fits, the 
diversification of modeling frameworks, and the movement toward FAIR data all point to 
the need for a unified descriptive standard for statistical models. \hs3 addresses this 
need by providing a framework-independent, human-readable, machine-validated 
serialization format capable of expressing the full complexity of modern likelihoods.

Having established the need for a standard, the following section discusses the design principles.

\section{Design Principles of \hs3}

The development of \hs3 is guided by a set of core design principles aimed at providing a 
unified, descriptive, and implementation-independent representation of statistical models 
in high-energy physics. These principles reflect both the practical needs that emerged 
from the evolution of the HEP software ecosystem and the conceptual requirements for 
long-term preservation, interoperability, and scientific transparency.

But aside from mathematical and technical considerations, our community would not be well-served by the addition of yet-another statistical model format. \hs3 has not been designed in isolation, though, but in close communication with \texttt{ROOT} and \texttt{pyhf} developers, and the standard was refined in a consultation process with experts from the LHC experiments. Anything that can be represented in \texttt{RooFit} and \texttt{pyhf} can be represented in \hs3, conversions for \texttt{RooFit} workspaces are in place and \texttt{ROOT} itself already has preliminary support for \hs3.

\subsection{Descriptive domain-specific language}

At its core, \hs3 is a descriptive domain-specific language (DSL) for statistical 
modeling. Rather than encoding the execution logic or algorithmic instructions, \hs3 
provides a declarative specification of the mathematical structure of a statistical 
model. This includes probability distributions, functions, parameters, datasets, and the 
relationships among them. \hs3 files unambiguously define the mathematical structure of statistical models, but do not prescribe which algorithms should be used to evaluate models, maximize likelihoods, sample posteriors, and similar; these decisions are left to software that implements \hs3 support.

This explicit separation between mathematical semantics and software implementation is an important feature of the design. By representing models declaratively rather than procedurally, \hs3 allows different software frameworks to consume, interpret, and evaluate the same model with consistent semantics, but individual algorithmic and computational approaches. This enables robust cross-validation and facilitates 
reproducibility, and at the same time allows users to select the tools and algorithms best suited for their use case.

\subsection{Completeness and interoperability}

A central goal of \hs3 is to be feature-complete with respect to the needs of the HEP community.  This includes support for both binned and unbinned data, composite distributions, auxiliary measurements, user-defined functions, and extended models. 

Here, \texttt{RooFit}, the most widely 
used and currently most expressive statistical modeling framework in collider physics, provides an excellent design baseline: \texttt{RooFit} has a long-proven track record and has evolved alongside and according to the needs of multiple high-energy physics experiments. Thus, being on par with \texttt{RooFit} model semantics essentially corresponds to covering the needs of the community. This means that any \texttt{RooFit} model representable in a {\texttt{RooWorkspace}} must be representable in \hs3 without loss of information.

On the other hand, the success of pyhf, which is much more narrow in scope and expressiveness than a {\texttt{RooWorkspace}}, in the community demonstrates the value of a readable and compact \texttt{JSON}, compared to the binary \texttt{ROOT} file format and C++ frontend that \texttt{RooFit} workspaces are tied to.

\hs3 builds on the success of both \texttt{RooFit} and pyhf with a \texttt{JSON}-based format that fully covers the semantics of both, keeps the readability of pyhf, but is at least as expressive and semantically complete as \texttt{RooFit} -- but unlike them, is not tied to a specific set of algorithms and programming language.

Beyond this, \hs3 is designed for interoperability 
with other modeling frameworks. The standard aims to provide a common representation to which models from diverse tool chains can be translated and from which they can be reconstructed. This includes:

\begin{itemize}
    \item \texttt{ROOT}/\texttt{RooFit} via the builtin \texttt{RooFit} HS3 library,

    \item HistFactory-based tools such as \texttt{pyhf} via explicit conversion of the \texttt{JSON} files,

    \item CMS \texttt{Combine} through translation of DataCards and internal \texttt{RooFit} 
          objects,

    \item Julia-based modeling and inference frameworks such as \texttt{BAT.jl}

    \item in the future, potentially also other standalone applications and custom modeling environments.
\end{itemize}

A new model standard will only be successful if models in major existing frameworks can be converted to and from it, so that there is a clear risk-free path to adoption for the community. \hs3 has been designed to provide this genuine interoperability with the HEP statistical ecosystem.

\subsection{Human- and machine-readability}

\hs3 is designed to be both machine-readable and human-readable. \texttt{JSON} is the canonical 
serialization format, chosen for its wide adoption, mature tooling, and ability to 
represent structured data in a clear and explicit manner. The extensive \texttt{JSON} ecosystem 
allows \hs3 models to be validated, manipulated, and version-controlled using standard 
tools across all major programming languages.

While \texttt{JSON} is the primary format, the structure of \hs3 is sufficiently generic to allow 
alternative serializations such as \texttt{YAML} or \texttt{TOML} for environments where human editing or 
configuration management is prioritized. Regardless of format, \hs3 enforces explicit 
naming of all objects—parameters, distributions, datasets, functions, and composite 
structures. References between objects are resolved by name rather than by position or 
syntactic scope. This makes \hs3 models robust against reordering, readable by domain 
experts, amenable to incremental modification, and version-control friendly.

\subsection{Models as computational graphs}

When \hs3 objects -- distributions, functions and data -- are combined, the computation of the resulting log-likelihood functions is an implicit directed acyclic graph (DAG). \hs3 itself does not prescribe a specific method of code execution, so implementations that support the standard are free to take advantage of the structure of this computational graph as far as their capabilities allow, e.g. in regard to parallel execution, visualization, performance optimization and so on. The same holds when using an \hs3 model for random (toy) data generation -- the only difference is that distribution objects are represented via their PDF for a likelihood evaluation, and via their random number generator for data generation. 

While \hs3 uses statistics language, not graph-theory language, the computational graph is, in fact, human-readable: \hs3 objects form the nodes of the graph and their dependencies form its edges.

A key design choice concerns the \emph{granularity} at which such graphs are expressed. 
In principle, any statistical model could be decomposed down to a minimal set of 
elementary operations (addition, multiplication, exponentiation, and so on) and 
represented as a fine-grained DAG. However, such a representation would be both 
impractical and counterproductive: it would produce models of enormous size, obscure the 
high-level mathematical structure, and prevent inference frameworks from taking 
advantage of high-level semantic information when optimizing computation. Also, JSON should not be abused as a generic programming language. For this reason, \hs3 
does not prescribe a unique or minimal decomposition of models into nodes. Instead, it 
allows \emph{multiple levels of abstraction} to coexist within the same graph.

This flexibility is a deliberate design feature. \hs3 supports both:
\begin{itemize}
    \item \textbf{high-level composite nodes}, such as \texttt{histfactory\_dist}, which 
          encapsulate the full bin-by-bin structure of a template-based model, and
    \item \textbf{low-level elementary nodes}, representing operations such as addition, multiplication, or analytic function evaluation.
\end{itemize}

This hybrid approach offers several advantages:
\begin{itemize}
    \item \textbf{Compositionality:} Users may construct models at whatever level of 
          abstraction best matches their analysis, mixing high-level and low-level 
          components freely.
    \item \textbf{Backend optimization:} Framework developers retain the freedom to 
          replace high-level nodes with efficient internal implementations (e.g.\ a 
          vectorized exponential or a backend-specific interpolation scheme), enabling 
          fast evaluation without altering the semantic content of the model.
    \item \textbf{Expressiveness:} Models that are conceptually representable in a given 
          framework may require decomposing or composing nodes when exchanged across 
          toolkits. \hs3 is designed to accommodate such transformations cleanly.
    \item \textbf{Transparency:} The structure of the model—including composite and 
          elementary components—is explicit and inspectable in the \hs3 representation.
\end{itemize}

The requirement that \hs3 graphs be acyclic ensures that model evaluation can map directly to a loop-free static single-assignment (SSA) form, which is easy to map to modern hardware platforms, including computing accelerators like GPUs.

\subsection{Versioning, governance, and FAIRness}

Long-term usability and stability require careful attention to versioning and governance. 
\hs3 adopts a semantic versioning scheme that distinguishes between:
\begin{itemize}
    \item structural changes to the standard,
    \item backward-compatible extensions,
    \item deprecated or removed constructs.
\end{itemize}
\hs3 models specify the standard version they adhere to, enabling software frameworks to 
validate compatibility and interpret constructs correctly.

The design of \hs3 is aligned with FAIR principles: models are intended to be 
\emph{findable}, through persistent identifiers and metadata; \emph{accessible}, through 
open formats; \emph{interoperable}, through clearly defined semantics; and 
\emph{reusable} through explicit structure and documentation. As the HEP community 
continues its transition towards open data and reproducible science, standards like \hs3 
are essential to ensure that statistical models remain usable over time, while software and hardware environments evolve or are replaced.

The long-term governance of \hs3—through a possible standards committee, RFC process \cite{rfc2119}, or 
release schedule—is discussed in more detail in the Outlook section. Although \hs3 is currently 
being developed collaboratively across experiments and software communities, future governance 
structures will help ensure stability, community participation, and broad adoption.

\section{Structure of the \hs3 Standard}

At the top level, an \hs3 
document consists of several sets of components: distributions, functions, 
data and likelihood definitions provide the fundamental components, while analysis prescriptions, parameter domains and metadata pre-define -- but do not limit -- ways of using them. In the following, we describe the structure and semantics of these components. A comprehensive description of the specifications of the HS3 standard, including the allowed components and the required syntax and naming conventions, can be found on the project webpage\footnote{\url{hep-statistics-serialization-standard.github.io}}.

\subsection{Top-level components overview}

An \hs3 file is organized into several top-level sections, each containing a list or 
mapping of objects indexed by unique names. The principal sections are:
\begin{itemize}
    \item \textbf{distributions:} Definitions of probability distributions, which may be 
          elementary (e.g.\ Gaussian, Poisson) or composite (constructed from other 
          distributions or functions, such as a mixture distribution).
    \item \textbf{functions:} Deterministic mathematical functions used to construct 
          distributions, transform parameters, or derive auxiliary quantities.
    \item \textbf{data:} Observed or simulated datasets, represented in a format 
          compatible with the corresponding distributions.
    \item \textbf{likelihoods:} Objects combining distributions and datasets into likelihoods, building statistical models, suitable for inference.
    \item \textbf{domains:} Definitions of parameter ranges, types (continuous, 
          discrete), and optional constraints.
    \item \textbf{parameter\_points:} Named sets of parameter values of special interest, 
          such as Standard Model predictions or benchmark points.
    \item \textbf{analyses:} High-level analysis prescriptions specifying parameters of 
          interest, nuisance parameters, likelihood components, and optional priors or 
          test-statistic definitions.
    \item \textbf{metadata:} Provenance information, authorship, references, 
          version numbers, and auxiliary descriptive fields.
    \item \textbf{misc:} Free-form implementation hints, plotting metadata, and 
          optional experiment-specific information not used in likelihood evaluation.
\end{itemize}
Every object in these sections is required to have a globally unique name, ensuring 
unambiguous references across the \hs3 document.

\subsection{Distributions and functions}

Distributions represent the probabilistic elements of a statistical model. \hs3 supports 
both \textbf{fundamental} and \textbf{composite} distributions:
\begin{itemize}
    \item \textbf{Fundamental distributions} include widely used analytic forms such as 
          Gaussian, Poisson, log-normal, and multivariate normal distributions. Their 
          parameters may be numeric constants or references to parameters, observables and functions.
    \item \textbf{Composite distributions} are constructed from other distributions or 
          functions. Examples include mixture models, constraint terms, and hierarchical 
          combinations of likelihood components.
\end{itemize}

\hs3 accommodates two important distinctions:
\begin{itemize}
    \item \textbf{Concrete vs.\ parameterized distributions:} A distribution may be fully 
          specified at serialization time, or may refer to symbolic parameters that are 
          assigned values only during inference.
    \item \textbf{Extended vs.\ non-extended distributions:} The standard distinguishes 
          between distributions that include an implicit event-count term (extended) and 
          those that describe only shapes (non-extended). This distinction is widely used 
          in \texttt{RooFit} and is essential for modeling Poisson processes.
\end{itemize}

High-level constructions such as \texttt{histfactory\_dist} encapsulate complex 
template-based likelihoods (multiple samples, modifiers, constraint terms) as single 
distribution nodes. This improves readability and allows frameworks to implement 
optimized evaluation strategies.

\subsection{Functions}

Functions in \hs3 represent deterministic mathematical transformations. They serve as 
building blocks for constructing composite distributions or deriving intermediate 
quantities. Function objects specify:
\begin{itemize}
    \item a \textbf{function type}, indicating whether the function is elementary 
          (e.g.\ addition, multiplication, exponentiation), user-defined, or 
          high-level (e.g.\ a HistFactory interpolation function);
    \item a list of \textbf{arguments}, which may include numeric constants, parameters, 
          other functions, or references to distributions.
\end{itemize}

For both functions and distributions, there exists a \texttt{generic\_function} or \texttt{generic\_distribution}, respectively. These support  simple algebraic transformations that do not warrant specialized high-level nodes. 

\subsection{Data}

The \textbf{data} section specifies the numerical observations to which likelihoods are 
fitted. \hs3 supports multiple data formats corresponding to the needs of different 
distributions:
\begin{itemize}
    \item binned data for HistFactory- or template-based models,
    \item unbinned event lists for analytic or continuous likelihoods,
    \item structured datasets referencing multiple observables.
\end{itemize}

\subsection{Domains and parameter points}

The \textbf{domains} section defines the allowed support and structure of parameters used by 
the model. Domains provide semantic information beyond simple numerical bounds, enabling 
inference tools to make informed choices about optimization and sampling strategies.

Each domain may specify lower and upper bounds for parameters. In the future, more properties of parameters might be supported, including
\begin{itemize}
    \item whether the parameter is continuous or discrete;
    \item optional transformations or natural parameterizations (e.g.\ log-scale for strictly positive parameters).
\end{itemize}

For example, a signal strength parameter constrained to positive values may be assigned a 
domain with lower bound zero and an associated log transformation, informing optimizers or 
samplers that exploration in log-space may be numerically preferable. Similarly, an integer-
valued nuisance parameter may be marked as discrete, indicating that continuous optimization 
methods are not appropriate.
Domains define the valid support and numerical structure of parameters, while priors remain 
separate model components that encode probabilistic assumptions over that support.

The \texttt{parameter\_points} section defines a list of points that can be composed of one or more parmeters with associated values. They can also be identified as constant or floating, allowing to use this section for a variety of applications, including
\begin{itemize}
    \item benchmark points,
    \item known minima or other special points of the distributions,
    \item suggested starting points for iterative processes such as MCMC sampling or minimization.
\end{itemize}

\subsection{Likelihoods}

A \textbf{likelihood} object in \hs3 specifies how distributions and datasets are combined 
to form a statistical model. A likelihood typically consists of:
\begin{itemize}
    \item one or more probability distributions,
    \item associated datasets.
\end{itemize}

\subsection{Analyses}

The \textbf{analyses} section defines high-level prescriptions for statistical inference. This includes specific hypothesis definitions for hypothesis tests and concrete interpretation frameworks for likelihoods in terms of parameters of interest and nuisance parameters with allowed domains and potential priors.
An analysis object may include:
\begin{itemize}
    \item a list of \textbf{parameters of interest (POIs)},
    \item the \textbf{nuisance parameters} that should be profiled or integrated out,
    \item the associated likelihood(s),
    \item optional \textbf{priors}, constraints, or penalty terms,
\end{itemize}
Analysis objects thus bridge the gap between the declarative structure of the model and 
actual inference tasks such as parameter estimation, hypothesis testing, or limit setting.

\subsection{Metadata and miscellaneous information}

The \textbf{metadata} section provides descriptive and provenance information intended to 
support transparency, reproducibility, and long-term preservation. Typical metadata 
fields include:
\begin{itemize}
    \item the \hs3 version,
        \item timestamps and software versions.
  \end{itemize}
  Future versions of the standard will also allow the metadata field to include
   \begin{itemize}     
    \item authorship and contact information,
    \item references to publications or DOIs,
    \item experiment or collaboration identifiers.
\end{itemize}
The metadata field is intended for formalized metadata that can be machine-readable in a standardized form.

The \textbf{misc} section on the other hand offers space for optional, implementation-specific, or 
contextual information that is not required for inference but may facilitate analysis 
workflows. Examples include:
\begin{itemize}
    \item plotting or visualization hints,
    \item recommended fit ranges,
    \item optimizer configuration,
    \item experiment-specific flags,
    \item annotations or comments.
\end{itemize}
These fields are explicitly non-normative: inference backends are free to ignore them, 
but they are preserved through serialization to support data annotation and auxiliary 
tooling. As a design principle, the values obtained from inference are not allowed to depend on the contents of the \texttt{misc} section in any way.

\section{Implementations and Integration with Existing Frameworks}

A central goal of \hs3 is to provide a descriptive and interoperable representation of 
statistical models that can be consumed and produced by the major modeling ecosystems 
used in high-energy physics. This section summarizes the existing \hs3 implementations, 
their relationships with established toolkits, and their current feature coverage. 
Collectively, these implementations demonstrate that \hs3 can serve as a common language 
for statistical modeling across \texttt{ROOT}, Python, Julia, and mixed workflows involving CMS 
\texttt{Combine}, \texttt{pyhf}, and \texttt{RooFit}.

\subsection{\texttt{ROOT} and \texttt{RooFit}}

The most mature \hs3 implementation is provided by \texttt{ROOT} through the 
builtin \texttt{RooFit} HS3 library, introduced in \texttt{ROOT}~6.36. This tool enables bidirectional 
conversion between {\texttt{RooWorkspace}} objects and \hs3 \texttt{JSON} representations. The 
conversion process maps \texttt{RooFit} objects -- distributions, functions, variables, datasets, 
and composite likelihoods - onto the corresponding \hs3 structures while preserving names, 
parameter configurations, normalization conventions, and default values.

Coverage of \texttt{RooFit} classes is extensive: the vast majority of standard 
\texttt{RooAbsPdf}, \texttt{RooAbsReal}, and \texttt{RooAbsArg} subclasses are supported, 
including both analytic and template-based PDFs, composite models, constraint terms, 
user-defined functions, and unbinned likelihoods. For most analyses, 
\textbf{round-trip compatibility} is achieved: converting a workspace to \hs3 and back 
yields an equivalent workspace.

Performance considerations are addressed through high-level composite nodes such as 
\texttt{histfactory\_dist}, which encapsulate complex template combinations as single 
objects rather than expanding them into thousands of elementary operations. This avoids 
excessively large \texttt{JSON} files and allows \texttt{RooFit} to employ optimized C++ implementations 
during likelihood evaluation.

The \texttt{RooFit} integration is the reference implementation for \hs3 semantics and serves as the 
compatibility baseline for other ecosystems.

\subsection{CMS Combine and DataCards}

CMS \texttt{Combine} is a powerful and widely used statistical inference framework built 
on \texttt{RooFit}. Its primary input format is the \texttt{DataCard}, a declarative text representation of 
template-based models augmented with custom \texttt{RooFit} classes, especially for the modeling of statistical uncertainties from the template generation (compare, e.g., Refs.~\cite{combine} and \cite{barlowbeeston}).  \texttt{DataCard}s can be parsed by \texttt{Combine} to 
    construct a \texttt{RooWorkspace}, which is then exported to \hs3 using the \texttt{ROOT} tools 
    described above.

Most CMS analyses employ custom \texttt{RooFit} classes and specialized modeling components that 
do not yet have representations within \hs3. Some of these are implementations of concepts that have later and separately been included in the public version of \texttt{RooFit}, such that conversion during export to \texttt{JSON} 
is an option. For others, the modular nature of \hs3 allows for 
the future inclusion of custom distribution types or function nodes that can represent 
CMS-specific modeling constructs explicitly.

\subsection{pyhf}

Models represented in \texttt{pyhf} \texttt{JSON} format can be systematically translated into \hs3 
by promoting each \texttt{pyhf} channel to a \texttt{histfactory\_dist}. The \hs3 representation retains all numerical information 
present in the original \texttt{pyhf} model, including shapes, modifiers, interpolation 
codes, and global parameters. Additional expressiveness in \hs3 allows such models to be 
embedded in larger statistical structures involving:
\begin{itemize}
    \item multiple likelihood components,
    \item auxiliary measurements,
    \item unbinned distributions,
    \item composite or hierarchical models.
\end{itemize}

Conversely, simple \hs3 models that fall strictly within the HistFactory paradigm can be converted into \texttt{pyhf} \texttt{JSON} with little effort, thanks to the intentional similarities in the syntax between pyhf's HistFactory \texttt{JSON} and \texttt{histfactory\_dist}. This provides a direct path for bridging \texttt{ROOT}-based 
analyses and pure-Python reinterpretation workflows.

\subsection{Python implementation: \texttt{pyhs3}}

The \texttt{pyhs3} package provides a pure-Python implementation of the \hs3 standard. Its 
design goals are simplicity, readability, and interoperability with Python-based 
statistical workflows. Core features include:
\begin{itemize}
    \item a schema-aware parser, validator, and emitter for \hs3 \texttt{JSON} files,
    \item internal representations of distributions, functions, domains, and likelihoods,
    \item conversion utilities for interfacing with \texttt{ROOT}, \texttt{pyhf}, and 
          \texttt{Combine}-generated models,
    \item tools for inspecting, modifying, and constructing \hs3 models in Python.
\end{itemize}

The Python implementation currently supports all standard distribution and function types 
commonly used in ATLAS and CMS analyses, including analytic PDFs, composite 
template-based models, and hierarchical constructions. Where possible, \texttt{pyhs3}
aims to preserve round-trip compatibility with \texttt{ROOT} by mirroring the expected structure 
of the computational graph.

Extensive validation tests ensure that \texttt{py\hs3} interprets \hs3 documents 
consistently with \texttt{ROOT}. These tests include parameter sweeps, numerical consistency 
checks, and comparisons of log-likelihood values on benchmark models.

\subsection{Julia implementation: \texttt{HS3.jl}}

The Julia implementation, \texttt{HS3.jl}, provides a performant and expressive interface 
to \hs3 models in the Julia ecosystem. The package offers:
\begin{itemize}
    \item a high-performance \texttt{JSON} parser tailored to the \hs3 schema,
    \item Julia-native data structures for distributions, functions, and likelihoods,
    \item compatibility with JuliaHEP modeling tools and inference packages,
    \item a foundation for future integration with high-performance likelihood evaluators.
\end{itemize}

While feature coverage is not yet complete at the level of \texttt{ROOT} or \texttt{py\hs3}, the 
Julia implementation already supports most commonly used analytic distributions and 
function types, with ongoing efforts to extend compatibility to template-based and 
HistFactory-style models. The flexibility of Julia's multiple dispatch and JIT 
compilation offers promising opportunities for fast likelihood evaluation, automatic 
differentiation, and gradient-based inference directly from \hs3 representations.

More broadly, \texttt{HS3.jl} demonstrates that \hs3 can serve as a common modeling 
language across languages and frameworks, enabling cross-validation and comparison of 
statistical models in heterogeneous environments.

\section{Applications}

Although \hs3 is a relatively young standard, it has already enabled several concrete use 
cases across high-energy physics. These early applications demonstrate the feasibility of 
publishing, validating, and reusing likelihood models in a fully descriptive format, while 
laying the foundation for more ambitious combined analyses and cross-domain workflows.

\subsection{Public likelihoods on HEPData}

HEPData now supports the publication of statistical models in \hs3 format through the \hs3 
badge system. The badge indicates that an analysis provides a complete, machine-readable 
likelihood description conforming to the \hs3 standard, alongside the usual tables, figures, 
and auxiliary files. Several ATLAS analyses have already been published using \hs3-based 
likelihoods, covering Higgs boson measurements \cite{ATLAS:2025VHWWh,ATLAS:2025hki}, searches for new physics \cite{ATLAS:2025TripleHiggs6b}, and  Standard Model analyses \cite{ATLAS:2023fourtop,ATLAS:2024HNLttbar,ATLAS:2025ttc,ATLAS:2024ttbarZratio136,ATLAS:2025ttyy,ATLAS:2024ttZ,ATLAS:2024ttW,ATLAS:2024ttgammaXsec,ATLAS:2024ttbar_pPb,ATLAS:2025SameSignTop,ATLAS:2023ttgammaAsym}.

\begin{figure}
    \centering\noindent
\includegraphics[width=\linewidth]{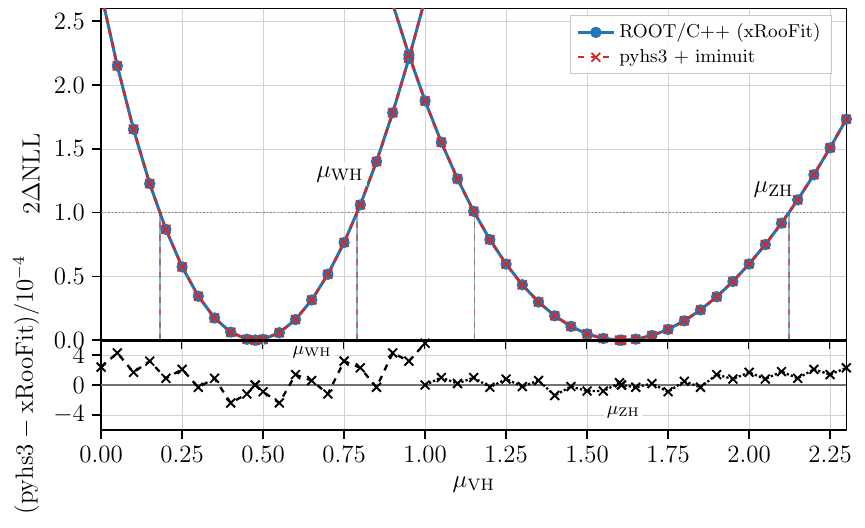}
    \caption{Comparison of profile likelihood curves, showing the distance to the global minimum of the negative logarithm of the likelihood against the signal strength parameters $\mu_{\textrm{WH}}$ and $\mu_{\textrm{ZH}}$, using the likelihood of a measurement of $VH\to WW$ of ATLAS data \cite{ATLAS:2025VHWWh}. Two independent implementations, the \texttt{C++ ROOT} implementation and the python package \texttt{pyhs3}, show good agreement for profile likelihood scans of both parameters of interest.}
    \label{fig:vhww}
\end{figure}

The impact of \hs3 on reproducibility is immediate: the descriptive nature of \hs3 allows 
external users to examine, reconstruct, and benchmark the published likelihoods without 
depending on specific software stacks such as \texttt{ROOT} or \texttt{pyhf}. 
For example, Figure \ref{fig:vhww} shows good agreement between these two independent implementations for a likelihood published by {ATLAS}.

Early adoption by the 
theory community through efforts such as the \texttt{spey-hs3} plugin has shown that \hs3 reduces the barrier to reinterpretation, enabling 
straightforward comparisons between theoretical predictions and the full statistical 
information of an analysis.

\subsection{Combined analyses and reinterpretations}

While large-scale combinations using \hs3 are still in their early stages, several proofs of 
concept have demonstrated the ability to combine multiple public likelihoods from different 
analyses into a single statistical model. Because \hs3 explicitly encodes structure, 
parameter names, constraints, and domains, such combinations can be performed without 
manual rewriting of models or ad-hoc reinterpretation of semantics.

Applications include:
\begin{itemize}
    \item re-evaluating published limits under modified theory predictions,
    \item exploring variations of effective field theory (EFT) hypotheses,
    \item comparing template-based results produced in \texttt{ROOT} and \texttt{pyhf} workflows,
    \item testing numerical stability and cross-framework consistency.
\end{itemize}

These examples illustrate the technical feasibility of \hs3-based combinations, and 
highlight the potential for future global fits using public models.

\subsection{Cross-domain adoption: the DEMOS project}

The DEMOS project aims to establish shared statistical modeling and data-preservation 
infrastructure across particle physics, astroparticle physics, nuclear physics, neutrino 
experiments, and related research fields. \hs3 has been selected as the reference format for 
statistical models within DEMOS, owing to its descriptive semantics, computational-graph 
structure, and ability to represent both binned and unbinned likelihoods.

Workflows under development in DEMOS include:
\begin{itemize}
    \item unified likelihood publication pipelines across multiple scientific domains,
    \item toolkits for cross-experiment reinterpretation,
    \item validation frameworks comparing models across \texttt{ROOT}, Python, and Julia backends,
    \item integration with long-term FAIR data preservation efforts.
\end{itemize}

The adoption of \hs3 within DEMOS demonstrates that the standard is not tied to collider 
physics specifically, but applies naturally to any domain employing likelihood-based 
inference.

\subsection{Outreach and teaching}

\hs3 is also well suited for outreach, tutorial material, and statistics education. Toy 
models involving simple Gaussian measurements, counting experiments, or template 
interpolations can be encoded in compact \hs3 documents that clearly expose their 
mathematical structure. Such examples are well-suited to be used in software tutorials, 
training schools, and courses introducing statistical inference in particle physics.  
Because \hs3 files are human-readable \texttt{JSON} documents, they serve as an intuitive bridge 
between conceptual teaching materials and full-fledged scientific analyses.

\section{Discussion and Outlook}

\subsection{Lessons learned: the central role of documentation}

One of the most important lessons from the development of \hs3 is the critical role of 
documentation in the adoption and sustainability of statistical modeling frameworks. The 
success of \texttt{pyhf} -- in particular its rapid uptake by both experimentalists and 
theorists -- demonstrated that high-quality documentation can be as important as technical 
capabilities. Clear examples, explicit modeling semantics, and readable formats foster 
trust and reduce barriers for new users.

As \hs3 itself is a standard, not a specific software package, these lessons on documentation carry over to the \hs3 specification: The standard is written in a human-readable form that is automatically rendered 
to Markdown, \texttt{HTML}, and \texttt{PDF} using a common documentation toolchain based on Pandoc.  
Examples, validation files, and diagrams are integrated into the repository to ensure that 
the semantics of each construct are unambiguous. This documentation-first approach has 
proved essential for community engagement and will remain a key aspect of \hs3 governance.

\subsection{Future extensions of the standard}

Several areas for future development have already been identified:
\begin{itemize}
    \item \textbf{Vector-valued and discrete parameters:} richer parameter types for 
          multivariate modeling and combinatorial hypotheses.
    \item \textbf{Advanced domain types:} periodic parameters, piecewise domains, and 
          constrained manifolds relevant for angular analyses or efficiency modeling.
    \item \textbf{Extended support for unbinned analyses:} including data 
          structures with complicated interrelated uncertainties.
\end{itemize}
As the statistical ecosystem evolves, \hs3 is designed to accommodate such extensions while 
maintaining backward compatibility.

\subsection{Governance and long-term maintenance}

Long-term sustainability of the standard will require a transparent governance model to guide 
its evolution while maintaining backwards compatibility and implementation stability. The 
precise governance structure is expected to be developed in consultation with the broader 
community and relevant stakeholders.

Possible governance mechanisms under consideration include:
\begin{itemize}
    \item a formal RFC (Request for Comments) process for proposing and discussing changes;
    \item versioned releases following semantic versioning principles;
    \item community review of feature requests and extensions;
    \item shared validation and compliance test suites across implementations.
\end{itemize}

Such mechanisms would provide a framework for coordinated development of the specification, 
allowing the standard to evolve in response to community needs while preserving stability for 
existing analyses and software implementations.

\subsection{\hs3 in the broader open-science landscape}

\hs3 is increasingly recognized as an enabling technology for FAIR likelihoods—models that 
are findable, accessible, interoperable, and reusable. The ability to describe statistical 
models independently of any specific software stack is essential for long-term data 
preservation at the LHC and beyond. 

More broadly, \hs3 offers a blueprint for cross-domain modeling standards. Its descriptive 
language, computational-graph structure, and explicit semantics make it applicable to a 
wide range of scientific disciplines that employ likelihood-based inference. As research 
communities continue their transition toward open data and reproducible science, \hs3 is 
well positioned to play a central role in the future of scientific modeling and 
publication. \hs3 aims to become for likelihoods and statistical models what \texttt{.root}-files have become for the HEP community: a durable, portable, and universally exchangeable representation.

\section*{Acknowledgements}

The authors acknowledge support by the Deutsche Forschungsgemeinschaft (DFG, German Research Foundation) through the project ``Public Likelihood Combination'' (project number 551818918) as well as under Germany’s Excellence Strategy  (project number 390833306, EXC 2121: Quantum Universe),
as well as the PUNCH4NFDI consortium supported by the DFG (project number 460248186).

This work has further received funding from the Bundesministerium für Forschung, Technologie und Raumfahrt (BMFTR, German Ministry for Research, Technology and Space) via the ErUM Data project ``DEMOS'' (FKZ 05D25GU6).

The authors gratefully acknowledge the computing time
provided on the Linux HPC cluster at Technical University
Dortmund (LiDO3), partially funded in the course of the
Large-Scale Equipment Initiative by the DFG (project 271512359).

We would like to thank all members of the HEP community for their intellectual contributions that facilitated the development of this work. This includes relevant groups such as the LHC reinterpretation forum and its members as well as other LHC working groups and their members, the statistics committees of both ATLAS and CMS and their members, the maintainers and developers of OpenData portals, specifically HEPdata. Further, we would like to thank a few individual experts such as Lukas Heinrich, Andy Buckley, and Wouter Verkerke for their valuable input, as well as all other contributors to and maintainers of the HEP statistical software ecosystem, especially Jonas Eschle, Tomas Dado, Massimiliano Galli and Robin Pelkner, and finally also the members of the theory community driving the adoption of the FAIR data standards, especially Sabine Kraml, Jack Araz and Martin Habedank.

\appendix

\bibliography{references}

@proceedings{workshop2k,
    editor = "James, F. and Perrin, Y. and Lyons, L.",
    title = "{Workshop on confidence limits, CERN, Geneva, Switzerland, 17-18 Jan 2000: Proceedings}",
    reportNumber = "CERN-2000-005",
    doi = "10.5170/CERN-2000-005",
    series = "CERN Yellow Reports: Conference Proceedings",
    month = "5",
    year = "2000"
}

@article{ROOT_NIMA_1997,
    author = "Brun, R. and Rademakers, F.",
    editor = "Werlen, M. and Perret-Gallix, D.",
    title = "{ROOT: An object oriented data analysis framework}",
    doi = "10.1016/S0168-9002(97)00048-X",
    journal = "Nucl. Instrum. Meth. A",
    volume = "389",
    pages = "81--86",
    year = "1997"
}

@article{roofit,
    author = "Verkerke, Wouter and Kirkby, David P.",
    editor = "Lyons, L. and Karagoz, Muge",
    title = "{The RooFit toolkit for data modeling}",
    eprint = "physics/0306116",
    archivePrefix = "arXiv",
    reportNumber = "CHEP-2003-MOLT007",
    journal = "eConf",
    volume = "C0303241",
    pages = "MOLT007",
    year = "2003"
}

@techreport{hf,
      author        = "Cranmer, Kyle and Lewis, George and Moneta, Lorenzo and
                       Shibata, Akira and Verkerke, Wouter",
      title         = "{HistFactory: A tool for creating statistical models for
                       use with RooFit and RooStats}",
      institution   = "New York U.",
      collaboration = "ROOT Collaboration",
      address       = "New York",
      reportNumber  = "CERN-OPEN-2012-016",
      month         = jan,
      year          = "2012",
      url           = "https://cds.cern.ch/record/1456844",
}

@software{pyhf,
author = {Heinrich, Lukas and Feickert, Matthew and Stark, Giordon},
doi = {10.5281/zenodo.1169739},
license = {Apache-2.0},
title = {pyhf},
url = {https://github.com/scikit-hep/pyhf/releases/tag/v0.7.0},
version = {0.7.0}
}

@article{pyhf-joss, doi = {10.21105/joss.02823}, url = {https://doi.org/10.21105/joss.02823}, year = {2021}, publisher = {The Open Journal}, volume = {6}, number = {58}, pages = {2823}, author = {Lukas Heinrich and Matthew Feickert and Giordon Stark and Kyle Cranmer}, title = {pyhf: pure-Python implementation of HistFactory statistical models}, journal = {Journal of Open Source Software} }

@article{barlowbeeston,
title = {Fitting using finite Monte Carlo samples},
journal = {Computer Physics Communications},
volume = {77},
number = {2},
pages = {219-228},
year = {1993},
issn = {0010-4655},
doi = {https://doi.org/10.1016/0010-4655(93)90005-W},
url = {https://www.sciencedirect.com/science/article/pii/001046559390005W},
author = {Roger Barlow and Christine Beeston}
}

@article{zfit,
title = {zfit: Scalable pythonic fitting},
journal = {SoftwareX},
volume = {11},
pages = {100508},
year = {2020},
issn = {2352-7110},
doi = {https://doi.org/10.1016/j.softx.2020.100508},
url = {https://www.sciencedirect.com/science/article/pii/S2352711019303851},
author = {Jonas Eschle and Albert {Puig Navarro} and Rafael {Silva Coutinho} and Nicola Serra}
}

@misc{rfc2119,
    series =    {Request for Comments},
    number =    2119,
    howpublished =  {RFC 2119},
    publisher = {RFC Editor},
    doi =       {10.17487/RFC2119},
    url =       {https://www.rfc-editor.org/info/rfc2119},
        author =    {Scott O. Bradner},
    title =     {{Key words for use in RFCs to Indicate Requirement Levels}},
    pagetotal = 3,
    year =      1997,
    month =     mar,
}

@misc{json,
    series =    {Information technology},
    number =    2017,
    howpublished =  {ISO/IEC 21778},
    publisher = {ISO},
    url =       {https://www.iso.org/standard/71616.html},
    title =     {{The JSON data interchange syntax}},
    year =      2017,
    month =     nov
}

@article{ATLAS:2024HNLttbar,
  author         = "{ATLAS Collaboration}",
  title          = "{Search for heavy right-handed Majorana neutrinos in the decay of top quarks produced in proton-proton collisions at $\sqrt{s}=13$ TeV with the ATLAS detector}",
  journal        = "Phys. Rev. D",
  volume         = "110",
  year           = "2024",
  pages          = "112004",
  doi            = "10.1103/PhysRevD.110.112004",
  eprint         = "2408.05000",
  archivePrefix  = "arXiv",
  primaryClass   = "hep-ex"
}

@article{ATLAS:2025VHWWh,
  author         = "{ATLAS Collaboration}",
  title          = "{Measurements of the production cross-sections of a Higgs boson in association with a vector boson and decaying into $WW^*$ with the ATLAS detector at $\sqrt{s}=13$ TeV}",
  journal        = "JHEP",
  volume         = "08",
  year           = "2025",
  pages          = "034",
  doi            = "10.1007/JHEP08(2025)034",
  eprint         = "2503.19420",
  archivePrefix  = "arXiv",
  primaryClass   = "hep-ex"
}

@article{ATLAS:2025ttc,
  author         = "{ATLAS Collaboration}",
  title          = "{Measurement of top-quark pair production in association with charm quarks in proton-proton collisions at $\sqrt{s}=13$ TeV with the ATLAS detector}",
  journal        = "Phys. Lett. B",
  volume         = "860",
  year           = "2025",
  pages          = "139177",
  doi            = "10.1016/j.physletb.2024.139177",
  eprint         = "2409.11305",
  archivePrefix  = "arXiv",
  primaryClass   = "hep-ex"
}

@article{ATLAS:2024ttbarZratio136,
  author         = "{ATLAS Collaboration}",
  title          = "{Measurement of the $t\bar{t}$ cross section and its ratio to the $Z$ production cross section using $pp$ collisions at $\sqrt{s}=13.6$ TeV with the ATLAS detector}",
  journal        = "Phys. Lett. B",
  volume         = "848",
  year           = "2024",
  pages          = "138376",
  doi            = "10.1016/j.physletb.2023.138376",
  eprint         = "2308.09529",
  archivePrefix  = "arXiv",
  primaryClass   = "hep-ex"
}

@article{ATLAS:2025ttyy,
  author         = "{ATLAS Collaboration}",
  title          = "{Observation of $t\bar{t}\gamma\gamma$ production at $\sqrt{s}=13$ TeV with the ATLAS detector}",
  journal        = "CERN-EP-2025-125",
  year           = "2025",
  eprint         = "2506.05018",
  archivePrefix  = "arXiv",
  primaryClass   = "hep-ex"
}

@article{ATLAS:2024ttZ,
  author         = "{ATLAS Collaboration}",
  title          = "{Inclusive and differential cross-section measurements of $t\bar{t}Z$ production in $pp$ collisions at $\sqrt{s}=13$ TeV with the ATLAS detector, including EFT and spin-correlation interpretations}",
  journal        = "JHEP",
  volume         = "07",
  year           = "2024",
  pages          = "163",
  doi            = "10.1007/JHEP07(2024)163",
  eprint         = "2312.04450",
  archivePrefix  = "arXiv",
  primaryClass   = "hep-ex"
}

@article{ATLAS:2024ttW,
  author         = "{ATLAS Collaboration}",
  title          = "{Inclusive and differential cross-section measurements of $t\bar{t}W$ production in $pp$ collisions at $\sqrt{s}=13$ TeV with the ATLAS detector}",
  journal        = "JHEP",
  volume         = "05",
  year           = "2024",
  pages          = "131",
  doi            = "10.1007/JHEP05(2024)131",
  eprint         = "2401.05299",
  archivePrefix  = "arXiv",
  primaryClass   = "hep-ex"
}

@article{ATLAS:2024ttgammaXsec,
  author         = "{ATLAS Collaboration}",
  title          = "{Measurements of inclusive and differential cross-sections of $t\bar{t}\gamma$ production in $pp$ collisions at $\sqrt{s}=13$ TeV with the ATLAS detector}",
  journal        = "JHEP",
  volume         = "10",
  year           = "2024",
  pages          = "191",
  doi            = "10.1007/JHEP10(2024)191",
  eprint         = "2403.09452",
  archivePrefix  = "arXiv",
  primaryClass   = "hep-ex"
}

@article{ATLAS:2024ttbar_pPb,
  author         = "{ATLAS Collaboration}",
  title          = "{Observation of $t\bar{t}$ production in the lepton+jets and dilepton channels in $p$+Pb collisions at $\sqrt{s_{\mathrm{NN}}}=8.16$ TeV with the ATLAS detector}",
  journal        = "JHEP",
  volume         = "11",
  year           = "2024",
  pages          = "101",
  doi            = "10.1007/JHEP11(2024)101",
  eprint         = "2405.05078",
  archivePrefix  = "arXiv",
  primaryClass   = "hep-ex"
}

@article{ATLAS:2025SameSignTop,
  author         = "{ATLAS Collaboration}",
  title          = "{Search for same-charge top-quark pair production in $pp$ collisions at $\sqrt{s}=13$ TeV with the ATLAS detector}",
  journal        = "JHEP",
  volume         = "02",
  year           = "2025",
  pages          = "084",
  doi            = "10.1007/JHEP02(2025)084",
  eprint         = "2409.14982",
  archivePrefix  = "arXiv",
  primaryClass   = "hep-ex"
}

@article{ATLAS:2025TripleHiggs6b,
  author         = "{ATLAS Collaboration}",
  title          = "{A search for triple Higgs boson production in the 6$b$ final state with the ATLAS detector}",
  journal        = "Phys. Rev. D",
  volume         = "111",
  year           = "2025",
  pages          = "032006",
  doi            = "10.1103/PhysRevD.111.032006",
  eprint         = "2411.02040",
  archivePrefix  = "arXiv",
  primaryClass   = "hep-ex"
}

@article{ATLAS:2023ttgammaAsym,
  author         = "{ATLAS Collaboration}",
  title          = "{Measurement of the charge asymmetry in top-quark pair production in association with a photon with the ATLAS experiment}",
  journal        = "Phys. Lett. B",
  volume         = "843",
  year           = "2023",
  pages          = "137848",
  doi            = "10.1016/j.physletb.2023.137848",
  eprint         = "2212.10552",
  archivePrefix  = "arXiv",
  primaryClass   = "hep-ex"
}

@article{ATLAS:2023fourtop,
  author         = "{ATLAS Collaboration}",
  title          = "{Search for four-top-quark production in the single-lepton channel at $\sqrt{s} = 13$ TeV with the ATLAS detector}",
  journal        = "JHEP",
  volume         = "06",
  year           = "2023",
  pages          = "063",
  doi            = "10.1007/JHEP06(2023)063",
  eprint         = "2304.01532",
  archivePrefix  = "arXiv",
  primaryClass   = "hep-ex"
}

@article{higgsmasscomb,
  author         = "{ATLAS \& CMS Collaborations}",
    title = "{Combined Measurement of the Higgs Boson Mass in $pp$ Collisions at $\sqrt{s}=7$ and 8 TeV with the ATLAS and CMS Experiments}",
    eprint = "1503.07589",
    archivePrefix = "arXiv",
    primaryClass = "hep-ex",
    reportNumber = "ATLAS-HIGG-2014-14, CMS-HIG-14-042, CERN-PH-EP-2015-075",
    doi = "10.1103/PhysRevLett.114.191803",
    journal = "Phys. Rev. Lett.",
    volume = "114",
    pages = "191803",
    year = "2015"
}

@article{higgscouplingcomb,
  author         = "{ATLAS \& CMS Collaborations}",
    title = "{Measurements of the Higgs boson production and decay rates and constraints on its couplings from a combined ATLAS and CMS analysis of the LHC pp collision data at $ \sqrt{s}=7 $ and 8 TeV}",
    eprint = "1606.02266",
    archivePrefix = "arXiv",
    primaryClass = "hep-ex",
    reportNumber = "CERN-EP-2016-100, ATLAS-HIGG-2015-07, CMS-HIG-15-002",
    doi = "10.1007/JHEP08(2016)045",
    journal = "JHEP",
    volume = "08",
    pages = "045",
    year = "2016"
}

@article{atlashiggs,
  author         = "{ATLAS Collaboration}",
    title = "{Observation of a new particle in the search for the Standard Model Higgs boson with the ATLAS detector at the LHC}",
    eprint = "1207.7214",
    archivePrefix = "arXiv",
    primaryClass = "hep-ex",
    reportNumber = "CERN-PH-EP-2012-218",
    doi = "10.1016/j.physletb.2012.08.020",
    journal = "Phys. Lett. B",
    volume = "716",
    pages = "1--29",
    year = "2012"
}

@article{cmshiggs,
  author         = "{CMS Collaboration}",
    title = "{Observation of a New Boson at a Mass of 125 GeV with the CMS Experiment at the LHC}",
    eprint = "1207.7235",
    archivePrefix = "arXiv",
    primaryClass = "hep-ex",
    reportNumber = "CMS-HIG-12-028, CERN-PH-EP-2012-220",
    doi = "10.1016/j.physletb.2012.08.021",
    journal = "Phys. Lett. B",
    volume = "716",
    pages = "30--61",
    year = "2012"
}

@article{combine,
  author         = "{CMS Collaboration}",
    title = "{The CMS Statistical Analysis and Combination Tool: Combine}",
    eprint = "2404.06614",
    archivePrefix = "arXiv",
    primaryClass = "physics.data-an",
    reportNumber = "CMS-CAT-23-001, CERN-EP-2024-078",
    doi = "10.1007/s41781-024-00121-4",
    journal = "Comput. Softw. Big Sci.",
    volume = "8",
    number = "1",
    pages = "19",
    year = "2024"
}

@software{trexfitter,
  author       = {Aly, Mohamed and
                  Dado, Tomas and
                  Held, Alexander and
                  Pinamonti, Michele and
                  Valery, Loic},
  title        = {TRExFitter},
  month        = feb,
  year         = 2025,
  publisher    = {Zenodo},
  version      = {0.9.7},
  doi          = {10.5281/zenodo.14845713},
  url          = {https://doi.org/10.5281/zenodo.14845713},
}

@article{BLUE,
    author = "Lista, Luca",
    editor = "Foka, Y. and Brambilla, N. and Kovalenko, V.",
    title = "{Combination of measurements and the BLUE method}",
    eprint = "1610.00422",
    archivePrefix = "arXiv",
    primaryClass = "physics.data-an",
    doi = "10.1051/epjconf/201713711006",
    journal = "EPJ Web Conf.",
    volume = "137",
    pages = "11006",
    year = "2017"
}

@article{bat.jl,
  author  = {Schulz, Oliver and Beaujean, Frederik and Caldwell, Allen and Grunwald, Cornelius and Hafych, Vasyl and Kr{\"o}ninger, Kevin and Cagnina, Salvatore La and R{\"o}hrig, Lars and Shtembari, Lolian},
  journal = {SN Computer Science},
  title   = {BAT.jl: A Julia-Based Tool for Bayesian Inference},
  year    = {2021},
  issn    = {2661-8907},
  month   = {Apr},
  number  = {3},
  pages   = {210},
  volume  = {2},
  day     = {12},
  doi     = {10.1007/s42979-021-00626-4},
  url     = {https://doi.org/10.1007/s42979-021-00626-4},
}

@article{juliahep-perf,
    author = "Stanitzki, Marcel and Strube, Jan",
    title = "{Performance of Julia for High Energy Physics Analyses}",
    eprint = "2003.11952",
    archivePrefix = "arXiv",
    primaryClass = "physics.comp-ph",
    reportNumber = "PNNL-SA-151985 DESY 20-056, DESY-20-056, PNNL-SA-151985",
    doi = "10.1007/s41781-021-00053-3",
    journal = "Comput. Softw. Big Sci.",
    volume = "5",
    number = "1",
    pages = "10",
    year = "2021"
}

@article{julia-in-hep,
    author = "Stewart, Graeme Andrew and others",
    title = "{Julia in HEP}",
    eprint = "2503.08184",
    archivePrefix = "arXiv",
    primaryClass = "hep-ex",
    doi = "10.1051/epjconf/202533701266",
    journal = "EPJ Web Conf.",
    volume = "337",
    pages = "01266",
    year = "2025"
}

@article{julia-in-hep-potential,
    author = "Eschle, Jonas and others",
    title = "{Potential of the Julia Programming Language for High Energy Physics Computing}",
    eprint = "2306.03675",
    archivePrefix = "arXiv",
    primaryClass = "hep-ph",
    doi = "10.1007/s41781-023-00104-x",
    journal = "Comput. Softw. Big Sci.",
    volume = "7",
    number = "1",
    pages = "10",
    year = "2023"
}

@article{cmslhcb,
    author = "CMS \& LHCb Collaborations",
    title = "{Observation of the rare $B^0_s\to\mu^+\mu^-$ decay from the combined analysis of CMS and LHCb data}",
    eprint = "1411.4413",
    archivePrefix = "arXiv",
    primaryClass = "hep-ex",
    reportNumber = "CERN-PH-EP-2014-220, CMS-BPH-13-007, LHCB-PAPER-2014-049",
    doi = "10.1038/nature14474",
    journal = "Nature",
    volume = "522",
    pages = "68--72",
    year = "2015"
}

@article{ATLAS:2025hki,
  author         = "{ATLAS Collaboration}",
    collaboration = "ATLAS",
    title = "{Measurements of Higgs boson production via gluon{\textendash}gluon fusion and vector-boson fusion using $H\rightarrow WW^*\rightarrow \ell \nu \ell \nu $ decays in pp collisions with the ATLAS detector and their effective field theory interpretations}",
    eprint = "2504.07686",
    archivePrefix = "arXiv",
    primaryClass = "hep-ex",
    reportNumber = "CERN-EP-2025-054",
    doi = "10.1140/epjc/s10052-025-14761-2",
    journal = "Eur. Phys. J. C",
    volume = "85",
    number = "12",
    pages = "1403",
    year = "2025"
}

\section{Minimal \hs3 Examples}

This appendix provides two minimal but illustrative examples of statistical models encoded 
in the \hs3 format. The first example demonstrates a simple univariate Gaussian likelihood 
with a single parameter of interest. The second shows a compact HistFactory-style 
construction using the high-level \texttt{histfactory\_dist} node.

\subsection{A simple univariate Gaussian model}

The model below describes a single Gaussian measurement of an observable~$x$ with known 
standard deviation $\sigma=1$ and an unknown mean parameter~$\mu$ serving as the parameter 
of interest. The \hs3 representation specifies the Gaussian distribution, the dataset 
containing the observed value, and an analysis object that links the two.

\begin{verbatim}
{
    "analyses": [
        {
            "domains": [
                "obs_parameters_of_interest"
            ],
            "likelihood": "obs",
            "name": "obs",
            "parameters_of_interest": [
                "mu"
            ]
        }
    ],
    "data": [
        {
            "axes": [
                {
                    "name": "x",
                    "value": 1.27
                }
            ],
            "entries": [
                [
                    1.27
                ]
            ],
            "name": "obs_gaussian_channel",
            "type": "unbinned"
        }
    ],
    "distributions": [
        {
            "mean": "mu",
            "name": "gauss_x",
            "sigma": "sigma",
            "type": "gaussian_dist",
            "x": "x"
        }
    ],
    "domains": [
        {
            "axes": [
                {
                    "max": 10.0,
                    "min": -10.0,
                    "name": "mu"
                }
            ],
            "name": "obs_parameters_of_interest",
            "type": "product_domain"
        },
        {
            "axes": [
                {
                    "max": 10.0,
                    "min": -10.0,
                    "name": "mu"
                },
                {
                    "max": 10.0,
                    "min": -10.0,
                    "name": "x"
                }
            ],
            "name": "default_domain",
            "type": "product_domain"
        }
    ],
    "likelihoods": [
        {
            "data": [
                "obs_gaussian_channel"
            ],
            "distributions": [
                "gauss_x"
            ],
            "name": "obs"
        }
    ],
    "metadata": {
        "hs3_version": "0.2",
        "packages": [
            {
                "name": "ROOT",
                "version": "6.38.04"
            }
        ]
    },
    "parameter_points": [
        {
            "name": "default_values",
            "parameters": [
                {
                    "name": "x",
                    "value": 1.27
                },
                {
                    "name": "mu",
                    "value": 0.0
                },
                {
                    "const": true,
                    "name": "sigma",
                    "value": 1.0
                }
            ]
        }
    ]
}
\end{verbatim}

This is the smallest practical \hs3 likelihood: it contains a distribution, a dataset, a 
domain for the parameter, a likelihood object, and an analysis prescription.

\subsection{A product of two Gaussian distributions}

The model below describes a product of two Gaussian measurements. The \hs3 representation contains the complete model, similar to the case above, but now including the product.

\begin{verbatim}
{
    "analyses": [
        {
            "domains": [
                "nuisance_parameters",
                "parameters_of_interest"
            ],
            "likelihood": "likelihood",
            "name": "my_analysis",
            "parameters_of_interest": [
                "mu1"
            ]
        }
    ],
    "data": [
        {
            "axes": [
                {
                    "max": 10.0,
                    "min": -10.0,
                    "name": "x",
                    "value": 1.8448742587493427
                }
            ],
            "entries": [
                [
                    -0.028567328469794262
                ],
                [
                    -0.09758959924367261
                ],
                [
                    0.8301414329794277
                ],
                [
                    -0.18001364208465098
                ],
                [
                    0.8853988033587967
                ],
                [
                    -0.2791754160017632
                ],
                [
                    1.168603380508273
                ],
                [
                    2.290388749097474
                ],
                [
                    0.18297688463530193
                ],
                [
                    1.8448742587493427
                ]
            ],
            "name": "toy",
            "type": "unbinned"
        }
    ],
    "distributions": [
        {
            "factors": [
                "g1",
                "g2"
            ],
            "name": "prod",
            "type": "product_dist"
        },
        {
            "mean": "mu1",
            "name": "g1",
            "sigma": "sigma1",
            "type": "gaussian_dist",
            "x": "x"
        },
        {
            "mean": "mu2",
            "name": "g2",
            "sigma": "sigma2",
            "type": "gaussian_dist",
            "x": "x"
        }
    ],
    "domains": [
        {
            "axes": [
                {
                    "max": 5.0,
                    "min": -5.0,
                    "name": "mu2"
                },
                {
                    "max": 10.0,
                    "min": 0.1,
                    "name": "sigma1"
                },
                {
                    "max": 10.0,
                    "min": 0.1,
                    "name": "sigma2"
                }
            ],
            "name": "nuisance_parameters",
            "type": "product_domain"
        },
        {
            "axes": [
                {
                    "max": 5.0,
                    "min": -5.0,
                    "name": "mu1"
                }
            ],
            "name": "parameters_of_interest",
            "type": "product_domain"
        },
        {
            "axes": [
                {
                    "max": 5.0,
                    "min": -5.0,
                    "name": "mu1"
                },
                {
                    "max": 5.0,
                    "min": -5.0,
                    "name": "mu2"
                },
                {
                    "max": 10.0,
                    "min": 0.1,
                    "name": "sigma1"
                },
                {
                    "max": 10.0,
                    "min": 0.1,
                    "name": "sigma2"
                },
                {
                    "max": 10.0,
                    "min": -10.0,
                    "name": "x"
                }
            ],
            "name": "default_domain",
            "type": "product_domain"
        }
    ],
    "likelihoods": [
        {
            "data": [
                "toy"
            ],
            "distributions": [
                "prod"
            ],
            "name": "likelihood"
        }
    ],
    "metadata": {
        "hs3_version": "0.2",
        "packages": [
            {
                "name": "ROOT",
                "version": "6.41.01"
            }
        ]
    },
    "parameter_points": [
        {
            "name": "default_values",
            "parameters": [
                {
                    "max": 10.0,
                    "min": -10.0,
                    "name": "x",
                    "value": 0.0
                },
                {
                    "max": 5.0,
                    "min": -5.0,
                    "name": "mu1",
                    "value": 0.0
                },
                {
                    "max": 10.0,
                    "min": 0.1,
                    "name": "sigma1",
                    "value": 1.0
                },
                {
                    "max": 5.0,
                    "min": -5.0,
                    "name": "mu2",
                    "value": 1.0
                },
                {
                    "max": 10.0,
                    "min": 0.1,
                    "name": "sigma2",
                    "value": 2.0
                }
            ]
        },
        {
            "name": "nominal_poi",
            "parameters": [
                {
                    "max": 5.0,
                    "min": -5.0,
                    "name": "mu1",
                    "value": 0.0
                }
            ]
        },
        {
            "name": "nominal_nuisance",
            "parameters": [
                {
                    "max": 5.0,
                    "min": -5.0,
                    "name": "mu2",
                    "value": 1.0
                },
                {
                    "max": 10.0,
                    "min": 0.1,
                    "name": "sigma1",
                    "value": 1.0
                },
                {
                    "max": 10.0,
                    "min": 0.1,
                    "name": "sigma2",
                    "value": 2.0
                }
            ]
        },
        {
            "name": "ModelConfig__snapshot",
            "parameters": [
                {
                    "max": 5.0,
                    "min": -5.0,
                    "name": "mu1",
                    "value": 0.0
                }
            ]
        }
    ]
}
\end{verbatim}

Figure \ref{fig:doublegauss-callgraph} contains the corresponding call graph of this model, as extracted from the python implementation pyhs3. 

\begin{figure}
    \centering
    \includegraphics[width=\linewidth]{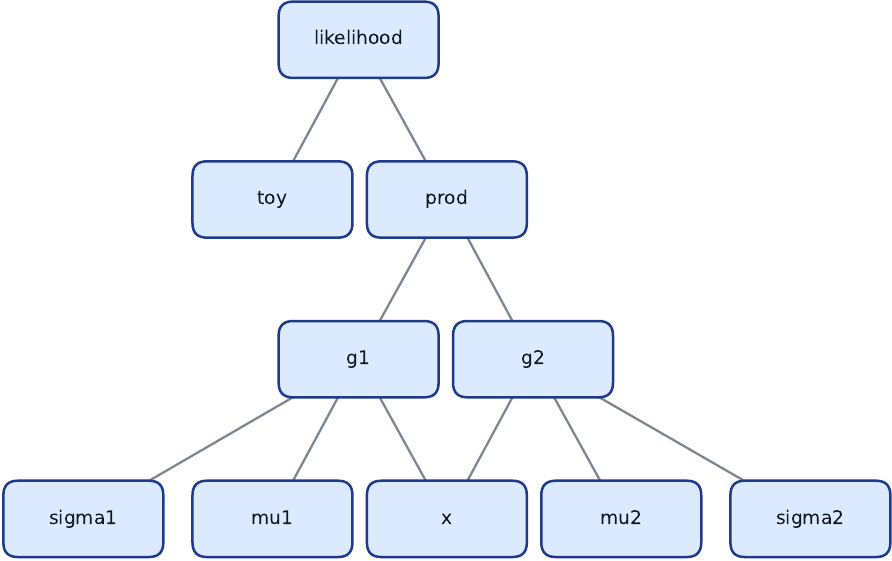}
    \caption{Dependency graph of a product of two Gaussian distributions encoded by the \hs3 layout}
    \label{fig:doublegauss-callgraph}
\end{figure}

\begin{figure*}
    \centering
    \includegraphics[width=\linewidth]{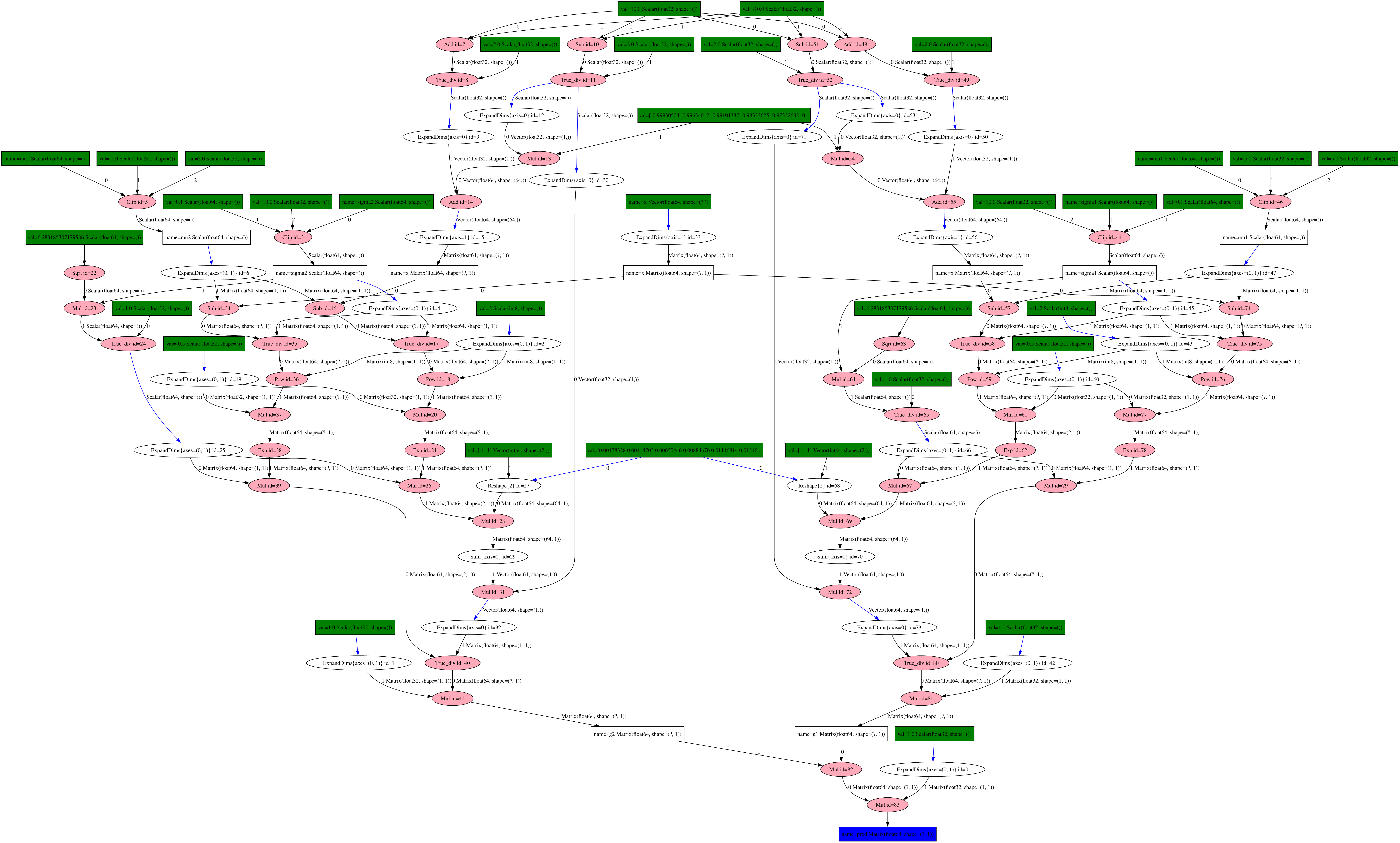}
    \caption{Call graph extracted from pyhs3 for a product of two Gaussian distributions}
    \label{fig:doublegauss-callgraph}
\end{figure*}

\subsection{A minimal HistFactory-style model}

The following example illustrates a compact, single-channel HistFactory model represented 
through the \texttt{histfactory\_dist} high-level node. The channel contains two samples, 
one signal and one background, both with template-based yields in three bins. A single 
normalization uncertainty affects the background.

\begin{verbatim}
{
    "analyses": [
        {
            "name": "main",
            "likelihood": "main_likelihood",
            "parameters_of_interest": ["mu"]
        }
    ],
    "likelihoods": [
        {
            "name": "main_likelihood",
            "distributions": ["model_channel1"],
            "data": ["observed_channel1"]
        }
    ],
    "distributions": [
        {
            "name": "model_channel1",
            "axes": [
                {
                    "name": "obs_x_channel1",
                    "max": 2.0,
                    "min": 1.0,
                    "nbins": 2
                }
            ],
            "samples": [
                {
                    "data": {
                        "contents": [20, 10]
                    },
                    "modifiers": [
                        {
                            "data": {
                                "hi": 1.05,
                                "lo": 0.95
                            },
                            "name": "syst1",
                            "type": "normsys"
                        },
                        {
                            "name": "mu",
                            "type": "normfactor"
                        }
                    ],
                    "name": "signal"
                },
                {
                    "data": {
                        "contents": [100, 0],
                        "errors": [5, 0]
                    },
                    "modifiers": [
                        {
                            "data": {
                                "hi": 1.05,
                                "lo": 0.95
                            },
                            "name": "syst2",
                            "type": "normsys"
                        },
                        {
                            "name": "mcstat",
                            "type": "staterror"
                        }
                    ],
                    "name": "background1"
                },
                {
                    "data": {
                        "contents": [0, 100],
                        "errors": [0, 10]
                    },
                    "modifiers": [
                        {
                            "data": {
                                "hi": 1.05,
                                "lo": 0.95
                            },
                            "name": "syst3",
                            "type": "normsys"
                        },
                        {
                            "name": "mcstat",
                            "type": "staterror"
                        }
                    ],
                    "name": "background2"
                }
            ],
            "type": "histfactory_dist"
        }
    ],
    "data": [
        {
            "name": "observed_channel1",
            "axes": [
                {
                    "name": "obs_x_channel1",
                    "nbins": 2,
                    "min": 1,
                    "max": 2
                }
            ],
            "contents": [122, 112],
            "type": "binned"
        }
    ],
    "metadata": {
        "hs3_version" : "0.1.90"
    }
}
\end{verbatim}

This example shows how complex template-based models can be represented cleanly using 
high-level \hs3 constructs without exposing the thousands of algebraic operations that 
underlie their numerical evaluation.

\end{document}